\documentclass{article}

\usepackage{cite,url,amssymb,amsmath,graphics,epsfig}
\usepackage{color}
\usepackage{enumerate}

\begin{document}

\title{Hilbert++  Manual}
\author{ Alessandro MIRONE\\
ESRF, BP 220, 38043 Grenoble, France}

\maketitle

\begin{abstract}
We present  here an installation guide, a hand-on mini-tutorial through  examples, and the theoretical foundations of  the Hilbert++ code.
\end{abstract}
\section{INSTALLATION}
The code is provided in both souce and binary forms.
The binaries are  meant to run under  the Linux-i386 system.
The distribution package  is composed of two archives.The first one  (let's call it  {\it installation archive}) which  provides all the necessary 
libraries that  Hilbert++ needs, and a second one, Hilbert++ itself, which  comprises sources, binaries, and examples.

Hilbert++  loads dynamically  the  libraries  of the  {\it installation archive} 
and does not depend on which Linux-packages are  installed   on your system 
 with the  exception of libc and libX11.
As these  libraries are quite stable one should be able to run Hilbert++ on any Linux system without recompiling it.

The current  {\it installation archive}  and the Hilbert++ packages can be retrieved from  these addresses :
 {\it http://ftp.esrf.fr/pub/scisoft/ESRF\_sw/scisoft\_ESRF\_sw\_linuxi386\_03.tar.gz }  and {\it http://ftp.esrf.fr/pub/scisoft/ESRF\_sw/linux\_i386\_03/hilbertxx.tar.gz} respectively.

The  installation archive must be untarred using  the following commands
\begin{verbatim}
cd /
tar -xzvf yourdownloaddirectory/scisoft_ESRF_sw_linuxi386_03.tar.gz
\end{verbatim}
this  creates the directory /scisoft/ESRF\_sw/linux\_i386\_03/.

The hilbertxx directory can be untarred from any directory with the command :
\begin{verbatim}
cd <whereever>
tar -xzvf yourdownloaddirectory/hilbertxx.tar.gz
\end{verbatim}

\section{ Hand on mini-tutorial by examples }

To launch the program you go to
hilbertxx/examples/exa\_{\it class}, where {\it class} corresponds to an  example class.
The example classes provided so far are : 
\begin{enumerate}[{Ex.} a)]
\item {\it class=}2p3d  for L2-L3 spectroscopy 
\item   {\it class=}rixs for 1s-3p RIXS spectroscopy, with a 3d resonant intermediate state,
   in transition metal oxides.

\item   {\it class=}df  for dipolar $d$ to $f$ transitions.

\end{enumerate}
You choose an example system by moving to one of the subdirectories. For instance
\begin{verbatim}
 cd hilbertxx/examples/exa_2p3d/Mn3+/
\end{verbatim}
Once you are in the exa\_{\it class}/{\it system} directory you start the application by running the Python script :
\begin{verbatim}
../runthisscript
\end{verbatim}

You get a command line  prompt:
\begin{verbatim}
>>>
\end{verbatim}
and you can get some help by typing
\begin{verbatim}
>>>help()
\end{verbatim}
The help function prints some brief instructions for every  Hilbert++-command (hxx-command).
A hxx-command is always one line long and writing it does not require any knowledge of Python.
 The advanced user can write Python scripts.
We explain  for each example the available commands

\subsection{The 2p3d examples class}
   \subsubsection{Mn3+}
cd to the appropriate directory 
\begin{verbatim}
cd hilbertxx/examples/exa_2p3d/Mn3+/
\end{verbatim}
and start the script
\begin{verbatim}
../runthisscript
\end{verbatim}

You get now a Python prompt. You can type several commands. In this manual we are going to explain all the commands.  The command {\it help()} which produces the following list of available commands :
{\scriptsize
\begin{verbatim}
 AVAILABLE HXX-COMMANDS ( examples of )
set(system)    # or use directly python syntax to change the attributes of object  system.
               # When using the set(system) function you can also give a filename of a previously
               # saved ( and edited ) file

save(system)   #  You are asked a filename and  you have to quote it like "filename"

scriviFiles(5,prefix="datas", nhopped=0)  # To generate matrices files. This command creates the Hilbert space
               # and writes the terms composing the Hamiltonian into (sub)directory "datas" which must exist.
               # The above example generates a Hilbert space where 5 electrons are distributed in the 3d
               # shell in all possible combinations.
               # Giving nhopped=n as argument, besides the configurations with 5 electrons on 3d also the configurations
               # with 5+i electrons on 3d and 10-i on Ligands orbitals are considered ( 0<=i<=nhopped )

system.case="datas" # This chooses the system that you have expanded with scriviFiles command.
               #    You can acces this property also with the set(system)    command

res=system.GetSpectrum() #   to calculate a spectrum.  res is a list.res[0] is an array of floats : the energies.
               # res[i>=1] are the resonances ( real and imaginary part). Dipolar transitions are calculated as
               # res[i], where i=1,2,3 and Mz= -1,0,1

res=system.GetSpectrum([ fm1,f0,f1 ]) # Absorption (it is res[1].imag)  for a  definite polarisation. 
               # Fm1,f0,f1 are three coefficients.
               # They can be complex. So you can define any polarisation :
               # For Z polarisation [ fm1,f0,f1 ] = [ 0, 1, 0]
               # For X polarisation [ fm1,f0,f1 ] = [-1/sqrt(2.0) , 0 ,1/sqrt(2.0)]
               # For Y polarisation [ fm1,f0,f1 ] = [ 1.0j /sqrt(2.0) , 0 ,1.0j/sqrt(2.0)]
               # ..... and so on

write(res,filename)   #  To save res on a file
               #  and the file will have several columns ( 1 for energies and two (real, imag) for each polarisation)

Plot(Curve(res[0], res[n>=1].imag, Pen(Red), "legend")) # To plot. Change n to select the polarisation that you want
               #  The plotting feature uses qt, iqt and qwt. If you run a long job in the background you must
               # desactivate this graphics feature by commenting out, in the file init.py, the lines where such packages
               # are imported

Es, S2s, L2s, SL2, occPs , Szs, Lzs= system.GetESLcounters()  #  to get statistics: a list of E, the list of 3d  L2 ,
               # the list of 3d S2, the list of 3d SL, a list of ligand expected occupation, a list of 3d Sz expectation

\end{verbatim}
}

The system is represented by the Python variable {\it system}. Such a variable is an object and  its properties are the 
 parameters. There are several ways to change such properties. One is to use 
the set(system) command. This command prints the list of parameters :

{\scriptsize
\begin{verbatim}
>>> set(system)
---- BASE HAMILTONIAN ---
  1) base_couche1_F0 :     5.0
  2) base_couche1_F2 :     12.4156828106
  3) base_couche1_F4 :     7.81967819912
  4) base_couche0_1_F0 :     5.5
  5) base_couche0_1_F2 :     6.86721502072
  6) base_couche0_1_G1 :     5.02109490016
  7) base_couche0_1_G3 :     2.85321756768
  8) base_SO_0       :     6.568603656
  9) base_SO_1       :     0.05238772
 10) base_Sop_Zero   :     1e-05
 11) base_Sop_Minus  :     0.0
 12) base_Sop_Plus   :     0.0
 13) base_counterDL  :     -4
---- EXCITED HAMILTONIAN ---
 14) exci_couche1_F0 :     5.0
 15) exci_couche1_F2 :     13.1769757147
 16) exci_couche1_F4 :     8.299507532
 17) exci_couche0_1_F0 :     5.5
 18) exci_couche0_1_F2 :     7.6574518
 19) exci_couche0_1_G1 :     5.77390099368
 20) exci_couche0_1_G3 :     3.28715525784
 21) exci_SO_0       :     6.845918392
 22) exci_SO_1       :     0.066403136
 23) exci_Sop_Zero   :     1e-05
 24) exci_Sop_Minus  :     0.0
 25) exci_Sop_Plus   :     0.0
 26) exci_counterDL  :     -4
---- CALCULATION PARAMETERS ---
 27) case            :     ./
 28) reduc_1         :     0.8
 29) reduc_0_1       :     0.8
 30) all1            :     0.1
 31) El2l3           :     700
 32) all2            :     0.1
 33) shift           :     0
 34) npunti          :     500
 35) dxleft          :     -0.1
 36) dxright         :     0.1
 37) temp            :     0.009
 38) erange          :     0.1
 39) tolefact        :     1e-06
 40) shift_invert    :     0
 41) nsearchedeigen  :     10
 42) NstepsTridiag   :     250
 43) Vs              :     2.0
 44) Vp              :     1.0
 45) VC0             :     0.2
 46) VC1             :     0.0
 47) DREF            :     1.0
 48) ALPHAVC         :     -3.0
 49) ALPHAVSP        :     -3.0
 50) BONDS           :     [[-1.0, 0, 0], [1.0, 0, 0], [0, -1.0, 0], [0, 1.0, 0], [0, 0, -1.0], [0, 0, 1.0]]
 51) factorhopexci   :     1.0
 52) facts_hop       :     None
--------------------------------------------------------------------
 select a value to change , 0 to stop, filename to read values

\end{verbatim}
}
To change a parameter, one enters the corresponding number and then the new value.
If, instead of a number, one enters a file-name like {\it ``paramfile''}  of a previously saved parameter file (the quotes are important),
then that file will be loaded. Such a file can be edited with any text editor.

All energies are given in $eV$ and all distances in Angstrom.
The parameters can be divided in three blocks. The first two are atomic Hamiltonian parameters
for the base and excited configurations respectively.
The first seven atomic parameters are Slater integrals : three (F0,F2,F4) for the {\it d-d} interaction (couche1), 
and four for the interaction between the {\it 2p} shell and the {\it 3d } one ( couche0\_1).
The parameter $SO_n$ is the spin-orbit interaction for the $n^{th}$ shell ( in this example 0 is{\it 2p } 
 and 1 means {\it 3d }. The one-particle energy of the orbitals is not given in input. Therefore
the absolute energy scale of the spectra is arbitrary.

The parameters {\it Sop\_Zero , Sop\_Minus, Sop\_Plus } are the three components of an external exchange field acting
on the {\it 3d } shell. They multiply the three components of the {\it 3d} spin operator $S_z,~S_{-1},~S_{+1}$. These factors can be real or complex.
Finally the parameter   {\it counterDL } is an energy which multiplies
the number of electrons on the Oxygen orbitals.

The block called {\it CALCULATION PARAMETERS} contains other interesting parameters. We start discussing the last ten : {\it Vs, Vp, VC0, VC1, DREF, 
ALPHAVC , ALPHAVSP, BONDS, factorhopexci, facts\_hop}. These parameters are relevant for the Hamiltonian. They describe hybridisation and crystal field.

The hybridisation of the {\it 3d } shell with the Oxygen {\it 2p } orbitals is described by the Slater-Koster parameters Vs and Vp  for 
$\sigma =3 {\tilde  z} ^{2}-r^2$ and $\pi=\tilde  x  \tilde z  , \tilde  y  \tilde z $ orbitals, where $\tilde z$ is aligned along the bond direction.
The hybridisation term is summed over all the bonds given by the {\it BONDS} variable. The Slater-Koster parameters are rescaled as $(R_{bond}/DREF)^{ALPHAVSP}$
where $R_{bond}$ is the bond length.

The parameters {\it VC0 } and {\it VC1} work in a similar way and describe the crystal field. They are the energy shifts for
$\sigma$ and $\pi$ orbitals, respectively, assuming that  $\tilde z$  is  aligned along the bond direction.
The scaling  factor is   $(R_{bond}/DREF)^{ALPHAVC}$.

The other parameters are described here :
\begin{enumerate}[{P)}]
\item {\it case } : the directory where the Hilbert space has been generated by the {\it scriviFiles } command.  
\item {\it reduc\_1 } : the usual Slater integral reduction factor for {\it 3d } shell 
\item {\it reduc\_0\_1 } :  reduction factor for $2p-3d$ Slater integrals. 
\item {\it  all1     }     :  Lorentzian broadening before El2l3
\item {\it El2l3  }   : an energy  between L2 and L3.         
\item {\it  all2  }   : broadening after El2l3        
\item {\it shift  }   :    an energy shift.
\item {\it npunti }     :    number of points in the spectrum.
\item {\it dxleft }     :  the spectrum is calculated from dxleft....
\item {\it  dxright }   :    .... to dxright
\item {\it  temp    }    :  when several ground state eigenvectors are calculated they are Boltzmann averaged with a Temperature=temp.
\item {\it  erange  }  :   the lowest energy ground-states are calculated which are within an energy range$=erange$. The smaller the $erange$, the faster will be the calculation since  less ground states are considered.
\item {\it  tolefact   }     :   The  ground states having Boltzmanian weight less than tolefact are neglected
\item {\it  shift\_invert  }  :     not used.
\item {\it  nsearchedeigen } :    The Lanczos diagonalisation of the ground state Hamiltonian searches for nsearchedeigen eigenvectors. ( the smaller this number the faster the calculation )
\item {\it NstepsTridiag  } : dimension of the tridiagonalised matrix used for calculating the spectra.
\end{enumerate}

When the program is started, the system parameters are initialised with default values. 
If two files named {\it out36\_base } and {\it out36\_exci }  are found in the working  directory,
 the  Slater integrals and the spin-orbit-coupling are searched for in such files.
These files are not mandatory. If they are not there you will have to input  the entries by hand.

The {\it out36 } files are the output as from the Cowan's Rcn program.
Hilbertxx comes with a GUI which allows to run Rcn and  generate the out36 file.
To run  the  GUI, from the working directory :
\begin{verbatim}
hilbertxx_directory/cowan/cowan in36 
\end{verbatim}

There is a short help for every RCN parameter. You can play with the parameters and create the out36 files that you need.

The examples are already provided with the related  out36 files, so you need to run RCN only for the  cases you want to study. 

To run a simulation you must first create  a directory with  the Hamiltonian components in the Hilbert space.
Let's call it HCD ( hamiltonian components directory ).
You obtain this by the command {\it scriviFiles }.
When  you  run {\it ../runthisscript }, and  your working directory 
has subdirectory name {\it datas }, write the command : 
 
\begin{verbatim}
scriviFiles(4,prefix="datas", nhopped=2)
\end{verbatim}
the first argument is the minimum occupation number for the 3d shell. Such command 
creates a Hilbert space  for the base system and another for the
excited system. The base space  spans all the configurations having 4, 5 and 6 electrons on the 3d shell.
The excited space spans those having 5,6 and 7 electron on 3d and 5 on the Mn 2p shell.
The number of electrons on the Oxygen 2p orbitals can be either 10 or  9 or  8 and the total number of electrons is conserved.

The program considers only those Oxygen  orbitals that are obtained projecting the 3d orbitals 
through the hopping operator. 
 
When the HCD has been populated  with files created by the above command,
you are ready to calculate a spectrum.
The HCD needs to be created only once, and then you can play with parameters.

In the example directory there is already a file with parameters which correspond to a case
with strong hybridisation. Such a file is named {\it paramn } and can be  loaded 
using the {\it set(system) } command.

To calculate a spectrum :

\begin{verbatim}
  res=system.GetSpectrum()
\end{verbatim}

and to  plot the isotropic absorption
\begin{verbatim}
Plot(Curve(res[0], (res[1]+res[2]+res[3]).imag, Pen(Red), "legend")) 
\end{verbatim}

Several expectation values can be calculated with the following commands
\begin{verbatim}
Es, S2s, L2s, SL2, occPs , Szs, Lzs= system.GetESLcounters()
\end{verbatim}

Each  variable Es, S2s, L2s, SL2, occPs , Szs, Lzs is a list of expectation values of a given operator.
There is an expectation value for each calculated ground eigenvector.
The operators are  energy, $S.S$, $L.L$, $2S.L$, the occupation of the ligand orbitals,
$S_z$, $L_z$.  The angular operators are restricted  to the 3d shell.

\subsection{The RIXS examples class}
  The directory hilbertxx/examples/exa\_rixs/ contains the code to calculate
the 1s-3p RIXS.
   \subsubsection{Mn3+}
As in the previous example, you run the command
\begin{verbatim}
../runthisscript
\end{verbatim} 
from the example  directory hilbertxx/examples/exa\_rixs/Mn3+.
You get a summary  of the available commands by entering {\it help()}.
You can calculate 1s-3d absorption spectra, as in the previous example,
and RIXS.
The absorption can be calculated either for the quadrupolar or for the dipolar interaction.
The dipolar absorption can be calculated for  centrosymmetric systems.
To do this the program calculates a dipolar transition to a virtual 4p state which is 
projected through an hybridisation operator onto the Oxygen orbitals. 
The command to get the dipolar absorption is the same as for the previous example :
\begin{verbatim}
res=system.GetSpectrum()
\end{verbatim}

The quadrupolar absorption can be calculated for a defined polarisation.
You define the polarisation giving its 5 components on a L=2 basis.

\begin{verbatim}
res=system.GetSpectrum([fm2,  fm1,f0,f1, fp2 ])
\end{verbatim}

 where fm2,  fm1,f0,f1, fp2 are the five polarisation coefficients in  the angular momentum representation.
 For example for  xy polarisation
 [fm2,  fm1,f0,f1, fp2 ]  = [ 1,0,0,0,-1  ]
 while for  calculating x2y2
   [fm2,  fm1,f0,f1, fp2 ] = [ 1,0,0,0,1   ].

The {\it scriviFiles } command has one more argument {\it spinfixed }. It defaults to zero ( no spin constraint ).
If it is set to one, the  Hilbert space is generated under the constraint of a fixed total spin $S_z$.
This has in general little effect on the absorption  spectrum  because the spin-orbit  coupling is small
in the $3d$ shell and the calculation is faster.

The use of the {\it set(system) } command is the same as in the previous example.
The difference is that there are some more parameters for the final state ( for RIXS) and some more for the 4p-Oxygen hybridisation which 
is used to calculate the dipolar 1s-3d transition. 
The latter parameters are {\it Dips , Dipp}, which are the Slater-Koster $\sigma$ and $\pi$ parameters for 4p-2p hybridisation,
and {\it ALPHADIPO }, which is the analogous of {\it ALPHAVSP}. 

A  parameter file is provided  for the study case of  Mn : {\it ``paramnH'' }
The calculation must be run in the Hilbert space created with the following command :
\begin{verbatim}
scriviFiles(4, prefix="datas", nhopped=2, spinfixed=1  )
\end{verbatim}
The studied ion is in a non-centrosymmetric position and we can calculate both  quadrupolar and dipolar absorption.
The  following commands 

\begin{verbatim}
resx2y2 = system.GetSpectrum(  [ 1.0 , 0.0 , 0 , 0 ,  1.0 ] )
resyz   = system.GetSpectrum(  [ 0.0 , 1.0 , 0 , 1.0 , 0 ] )
resdip=system.GetSpectrum()
\end{verbatim}

allow to compute the absorption spectra for $x^2-y^2$ , $y z$  and dipolar polarisation respectively.

You can compare the polarisation dependence of the quadrupolar absorption :
{
\begin{verbatim}
Plot(Curve(resx2y2[0], resx2y2[1].imag, Pen(Red), "x2y2"),
   Curve(resyz[0], resyz[1].imag, Pen(), "yz"))
\end{verbatim}
}
and you can plot a dipolar isotropic spectrum :
\begin{verbatim}
Plot(Curve(resdip[0], (resdip[1]+resdip[2]+resdip[3]).imag, Pen(Red), "dipole"))
\end{verbatim}

Now you can calculate a RIXS spectrum : 
{
\scriptsize
\begin{verbatim}
res= system.GetRIXS( polarisationIn=[0.0, 1.0,0.0, 1.0,0.0] ,polarisationOut=[1.0,0.0], ein=26.43,
   eout1=630, eout2=790, dout=0.1, gammain=0.2, gammaout=  [ 0.5 ,  20 , 1.0  ])
\end{verbatim}
}
where the energy is tuned to the first absorption peak.
The RIXS  calculation is performed  considering only one ground state, the lowest one regardless of the temperature and 
$erange$ parameter.

\subsection{The $df$ examples class}
  The directory hilbertxx/examples/exa\_df/ contains the code to
  calculate dipolar $d$ to $f$ transitions. This example considers a crystal
  field on the f shell but not hybridisation.
   \subsubsection{Ho}
     This example is very similar to the $pd$ case. The only difference
     is that there are three parameters for the crystal field and no hybridisation.
\section{Description of the code}

The low level objects are coded in C++ and wrapped in Python. 
The description of the system and the manipulation of basic objects are done in Python.

There are two levels of use of Hilbert++ : the developer level and the user level.

At the user level one uses a command line  interface which allows to manipulate the parameters
of an already defined system.

At the developer level one can create  new systems by doing some scriptic programming in Python.

The  Hilbert space is spanned by a set of second quantisation base vectors:

$$
e = c_{i}^{+{}} c _{k}  ^ {+{}} ~ .~  .~  . c _ { {l} }  ^ { {+{}} }  | 0  \rangle    
$$

The code represents a  base vector by a C++ object  $eo$. This object   contains two chains of bits :
$eo.val$ and $eo.signs$. The $i^{th}$  bit of $eo.val$  is equal to the occupancy 
of the $i^{th}$ one-particle state ( 1 or 0). The $eo.sign$ chain of bits is used to keep track of
fermionic statistics and  is the integral modulus 2 of $eo.val$.
If, for example, $c.val = 00010010....$ then $c.signs  = 00001110...$.

The creation and annihilation operators are represented under the form of one particle base vectors :
  they are represented by a base vector $co$ whose $co.val$  has only one bit set to one: i.e.
the bit corresponding to the  particle state.
A  $co$ creation operator  acts on a base state  in the following way :
first the ``AND'' bitwise operation is applied to   $co.val$ and $eo.val$. 
When the  one particle state is already occupied in the vector $eo$ the ``AND''
 operator gives a non-zero bit string and the result is set to zero.
Otherwise the result is a new vector $reso$ multiplicated by a $\pm 1$ factor.
The vector $reso$ is composed by a   $reso.val$ which is obtained by the  ``OR'' bitwise operation 
on $co.val$ and $eo.val$. The signs strings   $reso.signs$  is the result of the
 ``XOR'' operation on   $co.signs$ and $eo.signs$.
The multiplicative factor is $+1$ if $(co.val ~ AND ~ eo.signs) = 0$, otherwise it is $-1$
 
Starting  from these basic objects ( base vectors and creation/annihilation operators), 
the generic operators and vectors are constructed. 
An operator is in general a linear combination of creation/annihilation operators or their  products.
A vector is a linear combination of base vectors.

The generation of a Hilbert space  starts from  a seed basis, which is a basis formed by one or a few base vectors,
and a wanderer, i.e. an operator which, operating on the basis,  creates new basis vectors. 
The wanderer is applied several times  on the basis, making it grow until no new base vectors are found.

The ground state space  spans all the configurations having 4, 5 and 6 electrons on the 3d shell.
The excited state one spans those having 5,6 and 7 electron on 3d and 5 on the Mn 2p shell.
The number of electrons on the Oxygen 2p orbitals varies in order  to conserve the total number of  electrons.

The number of Oxygen orbitals is in principle $6 N_b$ where $N_b$ is the number of bonds. 
As in the above model Hamiltonian the Oxygen orbitals are degenerated,  one can arbitrarily 
chose a rotated basis  for the Oxygen one-particle wavefunctions.
The basis that we choose is such that 
the first $10$ basis vectors are obtained projecting the  $3d$ space on the oxygen orbitals
 through the hopping operator and applying a Grahm-Schmidt orthonormalization. The remaining vectors
span a space that is not related  to this problem and are neglected.

For an absorption study two Hilbert spaces  are expanded, one for the ground configurations, and another
for the excited ones.
The different components of the Hamiltonian are calculated separately and written  in files , in the form
of sparse matrices. 

This is done  writing 
a Hamiltonian component ( kinetic energy, electron-electron interactions, spin-orbit coupling, etc.... )
as a generic operator.  
This operator  is subsequently applied to every basis vector $i$ of the considered Hilbert space.
The result is a linear combination of a small number of basis vectors to which the vector $i$ is connected
through the operator. 

The sparse matrix, representing the operator, is a sequence of triples ($j$,$i$,$c_{ji}$)
where $i$ and $j$      are two connected basis vector indices and $c_{ji}$  is
the matrix element.
The total Hamiltonian is built as a linear combination of its components.
The dipolar and quadrupolar operators connecting the Hilbert spaces between each other
are represented as space matrices as well.

To calculate absorption and RIXS spectra one   needs to compute eigenvectors and the respective eigenvalues  :  
$$
H X_n  = \lambda_n X_n
$$
spectra  :
$$
\langle  X | \frac{  {1} } {  {\omega - H+ i \gamma} }  |  X \rangle 
$$
inversion
$$
   \frac{  {1} } {  {\omega - H+ i \gamma} }  |  X \rangle
$$

To do this, one just needs  to 
 apply  the  Lanczos algorithm and the  conjugate-gradient method, 
which are based on the simple 
multiplication
$$
  y = H x
$$

We have implemented the  Lanczos algorithm with thick restart
 from Kesheng Wu \& Horst Simon\cite{kesheng}.
The absorption is obtained  finding first  the ground eigenvectors $X_n$ via the Lanczos method.
The vector $ D X_n$, where $D$ is the dipolar ( or quadrupolar) operator  is 
then calculated in the excited Hilbert space.
This vector is taken as an initial vector at the zero-th step  of the Lanczos tridiagonalisation
procedure. A large enough number $N$ of steps is performed to generate the diagonal $\alpha_i$ ($0 \le i \le N$)
and off diagonal $\beta_j $    ($1 \le i \le N$) elements.
Te resonance amplitude is then given by the following continued fraction :
$$
 \frac {1} { \omega + i \gamma - \alpha_{{0} }  -  \frac{  \beta_{1}^ {2} } { \omega + i \gamma  - \alpha_{ 1 }
  -  \frac{ \beta_{ 2 }  ^ { 2 } } { .  .  . } } } \label{continued}
$$

The imaginary part of the above expression is proportional to the absorption.

The RIXS signal is calculated in a similar way and one takes at the first   step of the tridiagonalisation procedure,
the vector
$$
  D_{out} \frac{  {1} } {  {\omega_in - H+ i \gamma} }  |D_{in}  X_n \rangle
$$
where $D_{in}$ and $D_{out}$ are the dipolar ( or quadrupolar ) operators for the incoming and outgoing 
photons, respectively.

\section{Details on the hybridisation components of the Hamiltonian}

We consider a metal ion surrounded by Oxygen atoms.
The total Hamiltonian is :

\begin{align}
  H=H_{ato} +{}  \nonumber \\ 
  \sum_{b} { (t_{\sigma,b}  d^{+{}}_{3 {\tilde  z} ^{2}-r^2} p_{{\tilde z}} +  t_{\pi,b} (  d^{+{}}_{ {\tilde  x  \tilde z} } p_{{\tilde x}} + d^{+{}}_{ {\tilde  y  \tilde z} } p_{{\tilde y}}                 ) +C.C. )  } +{}
    \nonumber \\ 
  \epsilon_p \sum_{b} ( p_{{\tilde x}}^{+{}}p_{{\tilde x}} + p_{{\tilde y}}^{+{}}p_{{\tilde y}}+p_{{\tilde z}}^{+{}}p_{{\tilde z}} ) + {} \nonumber \\ 
  \sum_{b} (V_{\sigma,b}
                            d^{+{}}_{3 {\tilde  z} ^{2}-r^2}d_{3 {\tilde  z} ^{2}-r^2} 
                            +V_{\pi,b}  (d^{+{}}_{ {\tilde  x  \tilde z}} d_{ {\tilde  x  \tilde z}}+
                         d^{+{}}_{ {\tilde  y  \tilde z}}d_{ {\tilde  y  \tilde z}})     
                    )
\end{align}

In this expression $H_{ato}$ is the atomic Hamiltonian. The sum $\sum_{b}$ runs over the
 metal-oxygen bonds $b$ and  the $\tilde z$ axis is aligned along 
the bond direction. For d metals the hopping is expressed  as a function of  the Slater-Koster parameters $t_\sigma$, $t_\pi$
which are rescaled according to the bond length.
For the same metals the electrostatic crystal field is described in terms of  the  parameter $V_{\sigma}$  which is
the energy shift of the $3 {\tilde z}^2 - r^2 $ orbital, and of $V_{\pi}$ which is the energy shift 
of the ${\tilde x} {\tilde  z} $ and ${\tilde y} {\tilde  z}$ orbitals.
The crystal field parameters are rescaled accordingly.
The sum over the spin is omitted in the above formulae for conciseness.
As to the  df case there is one more  the crystal field parameter : in
 total there are three parameters VC0, for $z^3$, VC1 for $(5z^2-r^2)x$
 and  $(5z^2-r^2)y$, and finally VC2 for $zxy$ and $z(x^2-y^2)$.

\section{Acknowledgement}
The author has benefitted from preliminary discussions with Peter Krueger, now at
the Universite' de Dijon, about Lanczos tridiagonalisation.
The code, originally written  C++, has been wrapped in python with the help
of Mickael Profeta.

\section{Final comment}

The code is provided not only as a binary, ready to use executable, but also
with the sources. This because, analogously to what happens in
research, the only interesting code is the one that has not yet  been
written. Having the source will hence allow you to better adapt it to
your research topic.
The supplied examples can nonetheless be very useful and many things can
be learned working  with the parameters.
Like  everybody else, the author does not know what he will be doing in
the future : may be   expanding the code in some way  or doing
something completely different.
However if you have some interesting subject we can possibly  discuss
it together.
The distribution does not contain yet examples with  neighbouring metal ions like in \cite{mirone}.
Future versions will. 

% One of this way is using python syntax. For example
% {\it system.base_couche1_F0 }

\end{document}